\documentclass[letterpaper,arxiv]{dae-handout}
\graphicspath{{./}{./figures/}{../Figures/}{../../Figures/}{./images/}{./graphics/}}

\newcommand{\trversion}{v0.1}

\usepackage[T1]{fontenc}
\usepackage[utf8]{inputenc}
\usepackage[bitstream-charter]{mathdesign}

\usepackage{microtype}
\usepackage{amsmath,amssymb}
\usepackage{graphicx}
\usepackage{xcolor}
\usepackage{booktabs}
\usepackage{tikz}
\usetikzlibrary{calc}

\definecolor{linkblue}{RGB}{30, 80, 180}
\definecolor{linkgreen}{RGB}{30, 120, 60}
\definecolor{linkred}{HTML}{AE0000}
\usepackage{hyperref}
\hypersetup{
  colorlinks=true,
  urlcolor=linkblue,
  linkcolor=linkgreen,
  citecolor=linkgreen,
  filecolor=linkred,
}

\definecolor{catred}{RGB}{160, 50, 50}           
\definecolor{catbg}{RGB}{255, 248, 248}          
\definecolor{cattext}{RGB}{70, 25, 25}           

\usetikzlibrary{calc}

\newcommand{\categoryerror}[1][0pt]{%
  \marginnote[#1]{%
    \begin{tikzpicture}
      \node[
        draw=catred,
        line width=1.2pt,
        fill=catbg,
        rounded corners=3pt,
        inner sep=6pt,
        text width=\marginparwidth-16pt,
        align=left
      ] {%
        {\bfseries\color{catred}\footnotesize Category Mistake}\\[3pt]
        {\scriptsize\color{cattext}%
        Treating an \textbf{epistemic construct} (model, representation,
        abstraction) as if it were an \textbf{ontic entity} (physical
        reality).\\[4pt]
        \textit{The map is not the territory.}\\[3pt]
        \hangindent=0.8em\hangafter=1
        • The timestamp is not the event\\[1pt]
        \hangindent=0.8em\hangafter=1
        • The message is not a physical object\\[1pt]
        \hangindent=0.8em\hangafter=1
        • The wave function is not reality%
        }%
      };
    \end{tikzpicture}%
  }%
}

\providecolor{fitered}{RGB}{180, 40, 40}
\providecolor{fitobg}{RGB}{255, 245, 245}
\providecolor{fitotext}{RGB}{80, 20, 20}

\definecolor{fitoamber}{RGB}{180, 120, 40}
\definecolor{fitoamberbg}{RGB}{255, 250, 240}

\newcommand{\fitodiagnosticwide}{%
  \medskip
  \noindent
  \begin{tikzpicture}
    \node[
      draw=fitoamber,
      line width=1.5pt,
      fill=fitoamberbg,
      rounded corners=4pt,
      inner sep=10pt,
      text width=\textwidth-24pt,
      align=left
    ] {%
      {\bfseries\color{fitoamber}\small The FITO Diagnostic}\\[6pt]
      {\normalsize\color{fitotext}%
      If a system uses \textbf{clocks to establish truth}, relies on
      \textbf{timeouts for safety}, resolves conflict by
      \textbf{``latest wins,''} and cannot reverse a step
      \textbf{without a log}---then it is operating under
      \textbf{FITO thinking}.%
      }%
    };
  \end{tikzpicture}%
  \medskip
}

\setcitestyle{numbers,square}
\providecommand{\citenamefont}[1]{#1}

\newcommand{\wave}{\textsc{Wave}}
\newcommand{\ghost}{\textsc{ghOSt}}
\newcommand{\snap}{\textsc{Snap}}
\newcommand{\oae}{OAE}
\newcommand{\fito}{FITO}
\newcommand{\pcie}{PCIe}
\newcommand{\ipu}{IPU}

\newcommand{\smartnic}{SmartNIC}
\newcommand{\us}{$\mu$s}

\title[Wave and the FITO Category Mistake]{The Forward-In-Time-Only Assumption in SmartNIC Resource Management: A Critique of Wave and the Case for Bilateral Interaction}
\author[Paul Borrill]{Paul Borrill, D\AE D\AE LUS\quad---\quad\texttt{paul@daedaelus.com}}
\date{March 2026 \quad\textbar\quad \trversion}

\begin{document}
\maketitle


\begin{customabstract}
\noindent
The datacenter industry is converging on \smartnic{}-based resource
management.  \wave{} (Humphries et~al., ASPLOS~'25) demonstrates
the practical feasibility of offloading kernel thread scheduling,
memory management, and RPC stacks to the ARM cores of Intel's
Mount Evans Infrastructure Processing Unit (\ipu{}).  The engineering
is careful and the results are honest: without \wave{}'s \pcie{}
latency mitigations, offloaded workloads degrade by 350\%.

We argue that this 350\% degradation is not an engineering problem
to be optimized away but a \emph{diagnostic symptom} of a deeper
architectural issue: \wave{}'s communication model is
Forward-In-Time-Only (\fito{}).  Every interaction between host and
\smartnic{} is a unidirectional message---event forward, decision
back---creating a temporal vulnerability window in which decisions
can become stale before they are enforced.  \wave{}'s entire
optimization stack (write-combining page table entries, prestaging,
prefetching, atomic transaction abort) exists to hide or tolerate
this window.

We apply the \fito{} diagnostic to \wave{}'s architecture
systematically, identify the category mistake it inherits from
Lamport's happened-before and Shannon's channel model, and show
how Open Atomic Ethernet's bilateral swap primitive---implemented
on the same Intel \ipu{} hardware---dissolves the latency,
atomicity, and timeout problems without engineering around them.
The \smartnic{} is the right location for resource management;
what is missing is the right communication primitive at that
location.
\end{customabstract}

\section{Introduction}
\label{sec:intro}

Every major cloud provider has deployed \smartnic{} infrastructure.
Amazon's Nitro offloads hypervisor and security functions; NVIDIA's
BlueField and AMD's Pensando provide programmable ARM cores at the
network edge; Microsoft acquired Fungible (now Azure Boost) for the
same purpose.  Google's \wave{} framework~\sidecite{humphries2025wave}{Humphries et~al., ASPLOS~'25}
takes the next step: offloading operating system \emph{policy
decisions}---thread scheduling, memory tiering, RPC routing---to
the ARM cores of Intel's Mount Evans \ipu{}.

This is a significant architectural move.  \wave{} separates policy
from mechanism at the \pcie{} boundary, placing decision-making
logic on the \smartnic{} while the host kernel retains all
enforcement mechanisms.  The approach recovers 16~host cores from
memory management offload, 8~cores from RPC offload, and improves
virtual machine performance by 11.2\%.

But \wave{} also reveals a problem it does not name.  Without its
full suite of \pcie{} latency optimizations---write-combining and
write-through page table entries, prestaged decisions, prefetched
transaction state---offloaded workloads run \textbf{350\% slower}
than the optimized version.  The entire engineering effort is
devoted to hiding the round-trip latency of the \pcie{} interconnect
between host and \smartnic{}.

\categoryerror

We argue that this latency is not the fundamental obstacle.  The
fundamental obstacle is the \emph{communication model}.  \wave{}
inherits a Forward-In-Time-Only (\fito{}) information flow: events
depart the host, traverse \pcie{}, arrive at the \smartnic{}; decisions
depart the \smartnic{}, traverse \pcie{}, arrive at the host.
Each direction is a unidirectional message.  The host learns
whether its state has changed since the \smartnic{}'s observation
only \emph{after} the decision arrives---creating a temporal
vulnerability window that is the source of every optimization
\wave{} deploys.

This paper makes three contributions:

\begin{enumerate}
  \item We apply the \fito{} diagnostic~\sidecite{borrill2026distributed}{Borrill, FITO Category Mistake}
    to \wave{}'s architecture and show that it satisfies all four
    criteria for \fito{} thinking: clocks to establish truth,
    timeouts for safety, transaction abort as conflict resolution,
    and forward-recovery only.

  \item We identify the \emph{vulnerability window} between
    host observation and \smartnic{} decision enforcement as
    an instance of indefinite causal order at the microsecond
    scale---the same phenomenon that creates race conditions in
    high-frequency trading~\sidecite{borrill2026distributed}{Borrill, FITO Category Mistake}, file
    synchronization~\sidecite{borrill2026icloud}{Borrill, iCloud FITO}, and RDMA completion
    semantics.

  \item We show how Open Atomic Ethernet (\oae{})'s bilateral
    swap primitive, implemented on the same Intel \ipu{} hardware,
    eliminates the vulnerability window, the timeout-based failure
    detection, and the transaction abort mechanism---not by
    engineering around \pcie{} latency, but by using the round-trip
    as the synchronization primitive itself.
\end{enumerate}

\newpage
\section{Background: The \wave{} Architecture}
\label{sec:background}

\wave{} builds on three prior systems from the same research
group.  \ghost{}~\sidecite{humphries2021ghost}{Humphries et~al., SOSP~'21} demonstrated that
Linux kernel scheduling policy could be moved to userspace via
an atomic transaction API, without modifying applications.
Shinjuku~\sidecite{kaffes2019shinjuku}{Kaffes et~al., NSDI~'19} showed that centralized
preemptive scheduling at 5\,\us{} granularity improves tail
latency by 88\%.  \snap{}~\sidecite{marty2019snap}{Marty et~al., SOSP~'19} implemented
Google's production RPC stack as a microkernel-style userspace
networking system.

\wave{} extends \ghost{}'s architecture across the \pcie{}
boundary: the scheduling agent runs on the \smartnic{}'s ARM
cores instead of host userspace.  The host kernel retains all
fast-path enforcement---context switching, TLB shootdowns, page
table updates---while the \smartnic{} makes the policy decisions
that drive them.

\subsection{Communication Model}

\wave{} implements two unidirectional queues per offloaded
component:

\begin{description}
  \item[Host $\to$ \smartnic{}:] State updates and event
    notifications (thread blocked, page fault, RPC arrived).
    Backed by MMIO with write-combining (WC) page table entries
    to batch multiple messages per \pcie{} transaction.

  \item[\smartnic{} $\to$ Host:] Policy decisions (schedule
    thread~$X$ on core~$Y$, migrate page~$Z$ to tier~$T$).
    Backed by MMIO with write-through (WT) page table entries
    for caching, plus software prefetching.
\end{description}

For memory management, which requires moving entire page table
entry sets at 1+\,Gbps, DMA replaces MMIO as the transport
mechanism.

\subsection{The Transaction Model}

\wave{} inherits \ghost{}'s atomic transaction commit API.
Decisions are created via \texttt{TXN\_CREATE()}, committed
atomically via \texttt{TXNS\_COMMIT()}, and checked via
\texttt{POLL\_TXNS()}.  The critical property: if the target
resource has changed between the time the \smartnic{} observed
the host state and the time the host kernel attempts to enforce
the decision, the transaction \textbf{aborts without corrupting
kernel state}.

The paper states: ``This strong guarantee becomes even more
essential when the userspace system software that makes decisions
operates across a high-latency \pcie{} interconnect.''

\subsection{PCIe Latency Mitigation}

\wave{}'s primary engineering contribution is a systematic stack
of optimizations to hide \pcie{} round-trip latency:

\begin{description}
  \item[WC PTEs:] Batch host-to-\smartnic{} MMIO writes
    before flushing.
  \item[WT PTEs:] Cache \smartnic{}-to-host MMIO reads,
    eliminating redundant round-trips.
  \item[Prestaging:] Precompute decisions before the host
    requests them (one prestaged decision per core for
    scheduling; batched page migrations for memory management).
  \item[Prefetching:] Issue \texttt{PREFETCH\_TXNS()} before
    updating kernel state, hiding the 750\,ns MMIO read latency.
\end{description}

The measured latencies tell the story.  A host MMIO read costs
$\sim$750\,ns (full \pcie{} round-trip).  MSI-X delivery
end-to-end costs $\sim$1,600\,ns.  Without optimizations, context
switch overhead is 13.3\,\us{}; with all optimizations applied, it
drops to 3.3--4.0\,\us{}.  The complete decision loop---from
\smartnic{} decision write to MSI-X delivery---takes 426\,ns.

\subsection{What Was Offloaded}

\wave{} offloads three components:

\begin{enumerate}
  \item \textbf{Kernel thread scheduling} (\ghost{} agent):
    FIFO, Shinjuku preemptive, and GCE VM scheduling policies.
    Performance degradation: 1.1\% (FIFO).

  \item \textbf{Memory management} (TMLAD): Machine-learning-based
    memory tiering classifies pages as hot/cold and migrates
    between DRAM tiers.  Reduces RocksDB memory footprint by
    79\%, recovers \textbf{16 host cores}.

  \item \textbf{RPC stack} (\snap{}): Packet steering policy
    offloaded to \smartnic{}.  Recovers \textbf{8 host cores};
    VM turbo performance improves 11.2\%.
\end{enumerate}

\subsection{The 350\% Result}

Without \wave{}'s optimizations, RocksDB throughput drops from
895K~req/s to approximately 258K~req/s---a \textbf{350\% degradation}.
This is the paper's most honest and most important result.  It
proves that the \pcie{} latency problem is real and that the
mitigations are essential, not cosmetic.

It also proves something the paper does not say: the communication
model is \emph{fighting the physics of the interconnect}, and the
optimizations are engineering workarounds for a foundational
mismatch.

\section{The \fito{} Diagnostic Applied to \wave{}}
\label{sec:fito}

\fitodiagnosticwide

The \fito{} (Forward-In-Time-Only) category mistake occurs when
a system treats an epistemic construct---a model, timestamp, or
ordering convention---as if it were an ontic entity, a physical
fact about the world~\sidecite{borrill2026distributed}{Borrill, FITO Category Mistake}.  The \fito{}
diagnostic identifies four hallmarks of this mistake.  We apply
each to \wave{}.

\subsection{Criterion 1: Clocks to Establish Truth}

\wave{} uses sequence numbers on messages between host and
\smartnic{}.  The host kernel timestamps events; the \smartnic{}
processes them in order of receipt, which is assumed to reflect
causal order.  The transaction commit API uses these sequence
numbers to detect stale decisions.

The assumption is that if the \smartnic{} processes event~$N$
before event~$N+1$, the decision for~$N$ is causally valid at
commit time unless an explicit abort condition fires.  This is
the Lamport happened-before relation~\sidecite{lamport1978time}{Lamport, CACM 1978}
applied to \pcie{}: temporal ordering serves as a proxy for
causal validity.

But temporal ordering across a \pcie{} boundary is not causal
ordering.  The \pcie{} bus provides transaction ordering
guarantees (a protocol convention), not causal ordering guarantees
(which require the round-trip measurement that the Slowdown
Theorem demands~\sidecite{borrill2026flp}{Borrill, OAE vs.\ FLP}).

\subsection{Criterion 2: Timeouts for Safety}

\wave{}'s watchdog mechanism uses a $\sim$20\,ms timeout to
detect \smartnic{} agent failure.  If the agent does not produce
a scheduling decision within the window, the host kills and
restarts it.

This is the Fischer-Lynch-Paterson impossibility
result~\sidecite{fischer1985impossibility}{Fischer, Lynch \& Paterson, JACM 1985} in miniature.  \wave{}
cannot \emph{know} whether the \smartnic{} agent has failed, is
slow, or is processing a complex decision.  It can only
\emph{wait} and then \emph{guess}.  The 20\,ms timeout is
exactly the mechanism FLP shows is insufficient for consensus
in an asynchronous system: you can set the timeout conservatively
(risking unnecessary restarts) or aggressively (risking premature
kills), but you cannot eliminate the uncertainty.

Under bilateral interaction, this uncertainty vanishes.  If the
host and \smartnic{} are interacting through a reflective
acknowledgment protocol, the host knows immediately whether the
\smartnic{} is still participating.  There is no timeout, no
guessing, no false positives.

\subsection{Criterion 3: Transaction Abort as Conflict Resolution}

When a \smartnic{} decision arrives at the host and the target
state has changed (process exited, page freed), the transaction
aborts cleanly.  The agent receives fresh state and tries again.

This is not literally ``latest wins,'' but it is functionally
equivalent: the most recent host kernel state is always
authoritative, and stale decisions are silently discarded.  The
mechanism is a forward-only pipeline: observe state $\to$ compute
decision $\to$ attempt enforcement $\to$ succeed or discard.
There is no bilateral negotiation, no symmetric resolution, no
mutual commitment.

The paper frames clean abort as a feature---and it is, compared
to silent corruption.  But from the \fito{} diagnostic, it
reveals that every decision is \emph{retrospective}: it is about
a state that existed in the past, and its validity depends on that
state being unchanged at commit time.  This is the
time-of-check-to-time-of-use (TOCTOU) problem elevated to an
architectural principle.

\subsection{Criterion 4: Forward-Recovery Only}

Recovery after agent failure requires restarting from host kernel
state.  The \smartnic{} agent cannot undo a committed decision.
The host kernel is the ``source of truth''---all \smartnic{} state
is derived and can be reconstructed.

This is forward-recovery only: the system moves forward from the
last known good state.  There is no mechanism for rolling back to
a prior consistent state, no compensation for decisions that
committed successfully but were semantically wrong.

\subsection{Verdict: 4/4}

\wave{} satisfies all four \fito{} criteria.  This is not a
deficiency of the engineering---the engineering is excellent.
It is a consequence of the communication model.  Unidirectional
queues between host and \smartnic{} make \fito{} semantics
\emph{inevitable}: every message flows forward in time, every
response is a separate, independent message, and every failure
mode requires temporal reasoning (was the decision still valid
when it arrived?).

\section{The Atomicity Problem}
\label{sec:atomicity}

\wave{}'s atomic transaction model is its most interesting
feature---and its most revealing limitation.  The transaction
lifecycle exposes the vulnerability window that \fito{}
communication creates:

\begin{enumerate}
  \item Host sends events to \smartnic{} reflecting state
    at time $T_1$.
  \item \smartnic{} computes decision at time $T_2$, where
    $T_2 > T_1 + \Delta_{\text{\pcie{}}}$.
  \item \smartnic{} sends decision back; it arrives at time $T_3$.
  \item Host attempts atomic commit at time $T_3$.
  \item If host state has changed since $T_1$:
    \textbf{transaction aborts}.
\end{enumerate}

The interval $[T_1, T_3]$ is the \emph{vulnerability window}.
During this interval, the host state may change arbitrarily:
processes may exit, pages may be freed, new threads may arrive.
\wave{}'s contribution is making this window as small as possible
(426\,ns for the decision loop) and making the abort path clean.

But the \emph{existence} of the vulnerability window is the
fundamental issue.  \wave{} acknowledges that a decision made
at time $T_2$ based on state observed at time $T_1$ may be
invalid by time $T_3$.  This is indefinite causal order at the
microsecond scale: you cannot know whether the host state changed
during the \pcie{} round-trip without completing the round-trip.

\subsection{The Prestaging Paradox}

\wave{}'s prestaging optimization is particularly revealing.
To hide \pcie{} latency, the \smartnic{} precomputes decisions
\emph{before the host needs them}: one prestaged decision per
core for scheduling, batched migration decisions for memory
management.

This is speculative execution at the operating system level.
The \smartnic{} is making decisions about events that have not
yet occurred, based on patterns it has observed.  When the event
actually occurs, the prestaged decision is used if it matches;
otherwise, a fresh decision is computed at the cost of additional
latency.

Prestaging carries the same fundamental risk as hardware
speculative execution: the speculation can be wrong, and the
cost of wrong speculation (wasted \smartnic{} compute, added
latency on the miss path) is nonzero.  More fundamentally,
prestaging assumes that the future will resemble the past---which
is the \fito{} assumption encoded as an optimization strategy.

\subsection{The 350\% Number as Diagnostic Evidence}

\wave{}'s 350\% degradation without optimizations is not merely
a performance result---it is a \emph{measurement of the \fito{}
tax}.  The difference between 258K~req/s (unoptimized) and
895K~req/s (optimized) represents the cost of fighting the
communication model.

Every optimization in \wave{}'s stack---WC/WT PTEs, prestaging,
prefetching---is a mechanism for pretending the \pcie{} round-trip
latency is not there.  The engineering succeeds (the gap narrows
to 1.1--7.4\% vs.\ on-host), but the effort required is itself
evidence that the communication model is mismatched to the
problem.

Under bilateral interaction, the round-trip is not an obstacle
to be hidden.  It \emph{is} the synchronization primitive.  The
round-trip propagation time becomes the mechanism by which host
and \smartnic{} establish mutual knowledge of state, rather than
a source of stale-decision risk.

\section{The Bilateral Alternative}
\label{sec:bilateral}

Open Atomic Ethernet (\oae{})~\sidecite{borrill2026flp}{Borrill, OAE vs.\ FLP} provides
a fundamentally different communication primitive: the
\emph{bilateral swap}.  Instead of unidirectional messages (event
forward, decision back), both sides exchange state
\emph{simultaneously at the interaction boundary}.  The exchange
is atomic: both sides either commit or neither does.  There is no
vulnerability window because there is no temporal gap between
observation and action---the observation \emph{is} the action.

\subsection{Dissolving the Vulnerability Window}

In \wave{}'s model:
\begin{quote}
  Host observes state $\to$ sends message $\to$ \smartnic{}
  computes $\to$ sends decision $\to$ host checks validity
  $\to$ commit or abort.
\end{quote}

In the bilateral model:
\begin{quote}
  Host and \smartnic{} exchange state atomically at the \pcie{}
  boundary $\to$ both sides have consistent view $\to$ decision
  is valid by construction.
\end{quote}

The key difference is that the bilateral swap establishes
\emph{common knowledge} at the moment of interaction.  There is
no subsequent validity check needed because both sides have
committed to the exchanged state at the boundary.  This is
precisely the Slowdown Theorem's
requirement~\sidecite{borrill2026flp}{Borrill, OAE vs.\ FLP}: the round-trip measurement is
the \emph{minimum} interaction required to establish causal order,
and \oae{} uses exactly one round-trip---no more, no less.

\subsection{Eliminating the Timeout}

Under bilateral interaction, the host does not need a watchdog
timer to detect \smartnic{} failure.  The bilateral swap either
completes (both sides participated) or does not complete (one
side did not respond within the bounded-time interaction window).
The host knows \emph{immediately} whether the interaction
succeeded.

This eliminates the FLP impossibility problem for the
host-\smartnic{} pair: the bounded-time bilateral interaction
provides exactly the synchrony assumption that FLP's asynchronous
model lacks~\sidecite{fischer1985impossibility}{Fischer et~al., JACM 1985}.  The \smartnic{} is
not ``unreachable for an unknown duration''---it is either
participating in the current interaction or it is not, and the
host learns which within a single round-trip.

\subsection{Eliminating Transaction Abort}

If the bilateral swap establishes common knowledge at the
boundary, there is no need for a transaction abort mechanism.
The ``stale decision'' problem does not arise because the
\smartnic{}'s decision is based on state that is guaranteed to
be current at the moment of commitment---the swap itself
constitutes the commitment.

This does not mean the bilateral model is infallible.  It means
that failure modes are different and simpler: the swap either
completes or it does not.  There is no intermediate state where
a decision has been made but not yet validated.

\subsection{The Intel \ipu{} as Bilateral Substrate}

The Intel \ipu{} (Mount Evans~\citep{intel2021mountevans},
Mount Morgan E2200~\sidecite{intel2025mountmorgan}{Intel, Hot Chips 2021 \& 2025}) provides the
hardware substrate for both \wave{}'s \fito{} model and an \oae{}
bilateral model.  The same ARM cores, the same \pcie{} interface,
the same packet processing pipeline---what changes is the
communication primitive implemented in the \ipu{}'s firmware.

Mount Morgan (E2200), presented at Hot Chips~2025, upgrades to
24~ARM Neoverse N2 cores at 400\,Gbps with quad-channel
LPDDR5-6400, providing ample compute for bilateral protocol
processing.  Crucially, no new hardware is required.  The
bilateral swap is a \emph{firmware-level change} to the
existing \pcie{} communication path: instead of two unidirectional
queues, a single bilateral exchange protocol that uses the
\pcie{} round-trip as its synchronization boundary.

This positions the Intel \ipu{} not merely as an acceleration
device for offloaded OS policy, but as a \emph{correctness
infrastructure}---the distributed equivalent of the memory
management unit (MMU) for semantic transaction guarantees
across the datacenter.

\section{Implications}
\label{sec:implications}

\subsection{For Google's Fleet}

\wave{} is deployed on Google's production infrastructure, which
runs on Intel Mount Evans \ipu{}s.  Replacing \wave{}'s \fito{}
queues with bilateral \oae{} interactions would eliminate the
entire latency mitigation stack (WC/WT PTEs, prestaging,
prefetching) while providing \emph{stronger} correctness
guarantees.  The engineering simplification is significant:
\wave{}'s optimization stack represents substantial complexity
that exists solely to compensate for the communication model.

\subsection{For Apple and iCloud}

Apple stores more than 8~exabytes of iCloud data on Google Cloud
Platform, spending approximately \$7~billion annually on AWS and
Google Cloud combined.  Apple's iCloud file synchronization
failures---documented
elsewhere~\sidecite{borrill2026icloud}{Borrill, iCloud FITO}---are caused by \fito{}
semantics at the application layer: last-writer-wins conflict
resolution, timestamp-based ordering, silent conflict suppression.

But the \fito{} assumption is also embedded in the infrastructure
layer, in the \smartnic{} communication model that manages the
servers hosting Apple's data.  Fixing iCloud at the application
layer is insufficient if the infrastructure underneath enforces
\fito{} ordering.  \oae{} on Google's \ipu{} fleet would
propagate correct-by-construction transaction semantics upward
through the entire stack---from \smartnic{} firmware through
kernel scheduling through file synchronization to the end user.

\subsection{For AI/ML Training Infrastructure}

Large-scale AI training clusters (100K+ GPUs) require
deterministic synchronization barriers for gradient aggregation,
checkpoint atomicity, and firmware consistency.  The
timeout-based failure detection used by \wave{} and similar
systems introduces non-determinism that manifests as silent
training divergence, checkpoint inconsistency, and
mixed-firmware cluster states~\sidecite{borrill2026atomicity}{Borrill, AI/ML Atomicity}.

Bilateral interaction at the \ipu{} layer provides deterministic
failure detection without timeouts: either the synchronization
barrier completes bilaterally or it does not.  For training
workloads where a single stale gradient can propagate through
millions of parameter updates, the difference between ``timeout
and hope'' and ``deterministic commitment'' is the difference
between silent corruption and guaranteed correctness.

\subsection{For the SmartNIC Industry}

The broader lesson is that the communication model matters more
than the core count.  Mount Morgan's upgrade from 16 to 24~ARM
cores, and from 200 to 400\,Gbps, does not change the
fundamental \pcie{} round-trip constraint.  Faster cores compute
decisions sooner, but the vulnerability window persists as long
as the communication model is unidirectional.

The entire \smartnic{} industry---Intel, NVIDIA, AMD, Broadcom,
and their cloud customers---is converging on the right hardware
location for resource management.  What the industry has not yet
recognized is that the communication primitive at that location
determines whether the investment yields engineering complexity
(\fito{}) or architectural simplicity (bilateral interaction).

\section{Related Work}
\label{sec:related}

\textbf{In-network concurrency control.}
Eris~\sidecite{li2017eris}{Li et~al., SOSP~'17} moves transaction ordering into the
network fabric, achieving 3.6--35$\times$ throughput improvement
by eliminating coordination between servers.  Eris demonstrates
that the network is the right \emph{layer} for consistency
enforcement; \oae{} argues it is the right \emph{primitive}.

\textbf{\smartnic{}-accelerated transactions.}
Xenic~\sidecite{schuh2021xenic}{Schuh et~al., SOSP~'21} uses \smartnic{} cores for
distributed transaction coordination with optimistic concurrency
control, doubling throughput.  Like \wave{}, Xenic offloads
policy to the \smartnic{} but retains \fito{} communication
semantics.

\textbf{\smartnic{} offloading frameworks.}
iPipe~\sidecite{liu2019ipipe}{Liu et~al., SIGCOMM~'19} provides an actor-based model for
offloading to \smartnic{}s, demonstrating both the benefits and
the pitfalls (performance regressions under naive offload) that
\wave{} addresses.

\textbf{Policy/mechanism separation.}
The split between policy and mechanism that \wave{} implements
at the \pcie{} boundary has a long genealogy in operating systems
research, from Liedtke's L4 microkernel~\sidecite{liedtke1995microkernel}{Liedtke, SOSP~'95}
through Exokernel to \ghost{}.  \wave{}'s contribution is applying
this split across an interconnect boundary; our contribution is
observing that the interconnect's communication model then
determines the system's semantic guarantees.

\textbf{The \fito{} category mistake.}
The forward-in-time-only assumption is traced to Shannon's
channel model~\sidecite{shannon1948mathematical}{Shannon, Bell Sys.\ Tech.\ J.\ 1948} and Lamport's
happened-before relation~\citep{lamport1978time}, formalized as
a category mistake in~\citep{borrill2026distributed}, and shown to
underlie the FLP impossibility
result~\citep{fischer1985impossibility,borrill2026flp}, the CAP
theorem~\sidecite{brewer2012cap}{Brewer, IEEE Computer 2012}, and the Two Generals Problem.
The present paper extends this analysis to \smartnic{} resource
management.

\section{Conclusion}
\label{sec:conclusion}

\wave{} demonstrates, at Google production scale, that the
\smartnic{} is the right location for operating system resource
management.  The engineering is careful, the evaluation is honest,
and the results are genuine.  We do not dispute any of this.

What we dispute is the communication model.  \wave{}'s two
unidirectional queues---event forward, decision back---inherit
the \fito{} assumption from Shannon's channel model and Lamport's
happened-before relation.  This assumption creates the
vulnerability window, necessitates the timeout-based watchdog,
and requires the entire \pcie{} latency optimization stack that
constitutes \wave{}'s primary engineering contribution.

The bilateral swap primitive, implemented on the same Intel \ipu{}
hardware that \wave{} already uses, dissolves these problems
rather than engineering around them.  The round-trip that \wave{}
hides becomes the synchronization mechanism that \oae{} uses.
The timeout that \wave{} tunes becomes unnecessary.  The
transaction abort that \wave{} tolerates becomes impossible by
construction.

The datacenter industry is converging on \smartnic{}-based resource
management.  The question is no longer \emph{where} to manage
resources---Wave, Nitro, BlueField, and Pensando have answered
that---but \emph{how} to communicate at the management boundary.
We argue that the answer is bilateral interaction, not
unidirectional messaging, and that the Intel \ipu{} is the
canonical platform on which to demonstrate this.

\vspace{1em}
\noindent\rule{\textwidth}{0.4pt}
\vspace{0.5em}

{\small\noindent\textbf{Acknowledgments.}
The author thanks Edward A.~Lee (UC Berkeley) for foundational
discussions on determinism and the Slowdown Theorem, and
Anjali Singhai Jain (Intel) for insights into \ipu{} architecture.
Friday morning discussions with the Mulligan Stew group sharpened
the \fito{} analysis.
AI tools (Claude, Anthropic) assisted with literature review and
drafting.}


\end{document}